# Effects of swift heavy ions irradiation parameters on optical properties of muscovite mica*

ZHANG Sheng-Xia(张胜霞) [1,2;1]　LIU Jie(刘杰)[1;2]　ZENG Jian(曾健)[2]　SONG Yin(宋银) [1]　MO Dan (莫丹)[1]　YAO Hui-Jun (姚会军)[1]　DUAN Jing-Lai(段敬来)[1]　SUN You-Mei (孙友梅)[1]　Hou Ming-Dong (侯明东)[1]

1) (Materials Research Center, Institute of Modern Physics, CAS, Lanzhou 730000, China)

2) ( University of Chinese Academy of Sciences, Beijing 100049, China)

**Abstract:** Muscovite mica sheets with a thickness of 25 μm were irradiated by various kinds of swift heavy ions (Sn, Xe and Bi) in HIRFL. The fluences ranged from $1\times10^{10}$ ions/cm$^2$ to $8\times10^{11}$ ions/cm$^2$. The electronic energy loss $(dE/dx)_e$ was increased from 14.7 keV/nm to 31.2 keV/nm. The band gap and Urbach energy of pristine and irradiated mica were analyzed by ultraviolet-visible spectroscopy. Periodic fringes in long wave length of the absorption spectra caused by interference phenomenon, were disturbed as the $(dE/dx)_e$ increased. It was suggested that the chemical bonds between Tetrahedral-Octohedral-Tetrahedral (TOT) layers of mica were destroyed. Thus the smooth surface was cleaved after irradiation. The band gap was narrowed down with the increasing $(dE/dx)_e$ and fluences. The values of Urbach energy were increased as the $(dE/dx)_e$ and fluences gradually increased. It was indicated that the amount of defects and the proportion of amorphous structure were increased in mica irradiated under increased $(dE/dx)_e$ and fluences. Fluences took a distinctly important role in optical properties of mica.

**PAS:** 61.80.Jh, 61.82.Ms, 78.40.Ha

## 1. Introduction

Muscovite mica was a well-known mineral material with a layered structure. It was sensitive to irradiation damage. Mica was often used to investigate the solid state detector. In addition, the surface of mica was easy to cleave to obtain a smooth plane. It was commonly used as a substrate material [1, 2]. A series of research has been done on the effects of swift heavy ions (SHI) irradiation on mica. Latent tracks can be produced on mica by SHI irradiation observed by Price in nineteen sixties [3], and the latent tracks after chemical etching was used to produce the nano-channels. From then on, track etched-pores were applied to prepare a variety of different shapes of nanowires in different diameters [4, 5]. Over decades, muscovite mica has played an important part in research of track formations caused by swift heavy ions (SHIs) [6–9], clusters [10, 11] or highly charged ions (HCIs) [12].

Latent track was produced by the coupling of several complex processes, including amorphization, disturbing, and atomic displacement. Each of the processes was the properties of the material, and its crystallographic orientation with respect to the incident ion beam [13]. As a SHI penetrated into the material surface, it would convert its energy to the atoms and electrons around its path. Plenty of atomic displacement was generated due to the electronic and atomic collision-cascade. Damage in track core caused by SHIs irradiation was mainly amorphous regions. Several theories had been modeled to describe the formation and size of latent tracks. The thermal spike model had successfully predicted the track radii of several materials. Modification of structure can lead to significant influence on properties of materials. A large number of studies have been focused on the band gap of the irradiated materials [16–19].

A lot of works has been done on the etching properties and latent tracks morphology or size of muscovite mica after swift heavy ions irradiation. Sukhnandan Kaur et al investigated the optical properties of natural phlogopite mica after irradiated by 80 MeV oxygen ions. Early in 1951, Popper analyzed the absorption coefficient for natural muscovite [14]. Dahr et al observed transmittance spectral of Indian muscovite mica as the wavelength ranges from 300 to 1000 nm [15]. However, the work on the optical properties of muscovite mica irradiated by swift heavy ions was not available at present. Thus the aim of our work was to get insight into the effects of swift heavy ions irradiation on the structural modification and optical properties of muscovite mica. A series of mica sheets were irradiated under different electrical stopping loss (dE/dx)$_e$ and ions fluences. The band gap and Urbach energy were calculated according to the UV-VIS spectra. The relationship of irradiation parameters and optical properties was discussed.

## 2. Experimental setup

Muscovite mica (KAl$_2$[AlSi$_3$O$_{10}$] (OH)$_2$) sheets with a thickness of 25 μm were irradiated by various kinds of swift heavy ions accelerated by Lanzhou Heavy Ion Accelerator (HIRFL). Irradiation parameters including ionic energy, fluences and the electronic energy loss (d$E$/d$x$)$_e$ were shown in Table 1. Prior to irradiation, mica sheets was cut into small pieces about 5 × 5 mm, and fixed on a copper substrate. Aluminum foils with different thickness were fixed in front of mica sheets to adjust the electronic energy loss (d$E$/d$x$)$_e$. The maximum penetration depth and the (d$E$/d$x$)$_e$ were calculated by the Stopping and Range of Ions in Matter (SRIM)-2008 software.

The ultraviolet- visible (UV-VIS, Lambda900, PE, Germany) spectroscopy in the transmission was employed in this experiment to get information about the optical properties of both the pristine and irradiated mica. The analytical wavelength was in the range of 200-1000 nm with a resolution of 1 nm. The direction of the incident lights was perpendicular to the surface of the samples. The transmission spectra were transferred into absorption spectra. The direct band gap, indirect band gap and the Urbach energy were obtained according to the spectra.

Table1 Detailed irradiation parameters applied in the experiment.

| Ions | Energy/(MeV) | (dE/dx)$_n$/(keV/nm) | (dE/dx)$_e$/(keV/nm) | Penetrate depth /(μm) | Fluence/(ions/cm$^2$) |
|---|---|---|---|---|---|
| Sn | 250 | 8.4×10$^{-2}$ | 14.7 | 12.3 | 1×10$^{11}$ |
| Xe | 2020 | 9.6×10$^{-3}$ | 15.8 | 102.4 | 1×10$^{11}$ |
| Xe | 1500 | 1×10$^{-2}$ | 18 | 79.5 | 1×10$^{10}$ |
| Xe | 1500 | 1×10$^{-2}$ | 18 | 79.5 | 1×10$^{11}$ |
| Xe | 1500 | 1×10$^{-2}$ | 18 | 79.5 | 8×10$^{11}$ |
| Bi | 1390 | 4.3×10$^{-2}$ | 30.9 | 53.6 | 1×10$^{11}$ |
| Bi | 1250 | 4.7×10$^{-2}$ | 31.2 | 49.1 | 1×10$^{11}$ |

## 3. Results and discussion

### 3.1 Surface cleavage

After irradiation, the optical properties of pristine and irradiated mica were analyzed by UV-VIS. The absorption spectra were presented in Fig.1. A shift towards the visible region in absorption spectrum was observed for irradiated mica according to Fig.1 (a). The optical absorbance increased with the increasing of electronic energy loss (d$E$/d$x$)$_e$. It was suggested that the bond cleavage and reconstruction could be generated in irradiated mica. For pristine mica, regular sine wave appeared in the long wavelength because of the interference effect of the light. Since the surface of pristine mica was molecularly smooth, it was beneficial to capture the interference phenomenon. As observed in Fig.1 (b), the regular curves were disturbed as the (dE/dx)$_e$ value increased from 14.7 to 31.2 keV/nm.

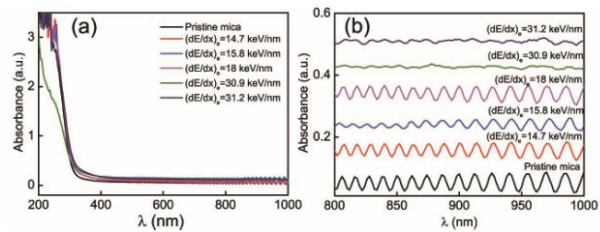

Fig.1 Ultraviolet absorption spectrums of muscovite mica. (a) Ultraviolet absorption spectrums of muscovite mica before and after irradiated under different (dE/dx)$_e$ and fluences; (b) variation of spectrums at long wavelength.

It was indicated that crystal structure of mica was modified due to irradiation, and the condition for interference was not satisfied. Muscovite mica was a two dimensional layered silicate mineral as shown in Fig.2. Concluded from the chemical formula of muscovite, one-fourth of Si was replaced by Al forming the aluminum silicate. Another part of the Al were bonded to O or OH forming octahedral structure, and a single octahedral was sandwiched between two identical tetrahedral (Si, Al)$_2$O$_5$. Trilayer aluminosilicate sheets were not electrically neutral, and a layer of K was located between the trilayer aluminosilicate to balance the charge. It was liable to cleave mica as the ionic bonding between the K layers and the trilayer aluminosilicate was weak[20]. As the swift heavy ions penetrated into the samples, the energy would convert to atoms and electrons around the path via ionization and electronic excitation processes. As the (dE/dx)$_e$ value was lower, such as 14.7 or 15.8 keV/nm, a small quantity of chemical bonds were destroyed. Flat face of several square centimeters could be emerged, and the larger areas of even surface were remained. The interference phenomenon was observed in the spectra without broad fluctuation. As (dE/dx)$_e$ increased to 30.9 or 31.1 keV/nm, a large amount of bonds between K and the trilayer aluminosilicate sheets were destroyed. The perfect layered structure ceased to exist. Regular wave were disordered, as seen in Fig.1 (b).

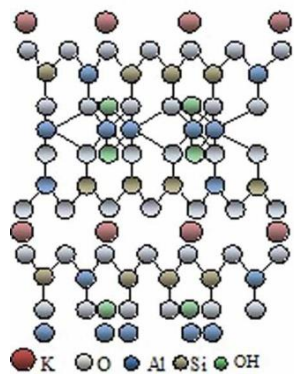

Fig. 2 Atomic-scale map of Muscovite mica

### 3.2 Effects of (dE/dx)$_e$ and fluences on optical properties of mica

The UV-VIS absorption spectra could be employed to investigate the structure transitions and the defects information in crystalline and non-crystalline materials. The optical band gap of mica could be calculated using Tauc's expression [21] which relates the absorption coefficient α and the incident photon energy $h\nu$ by the following equation:

$$\alpha(h\nu) = \frac{B(h\nu - E_g)^n}{h\nu} \quad (1)$$

where B was the correction coefficient. $E_g$ was the energy gap of the material. When n=1/2 and 2, equation (1) stands for the direct band gap and indirect band gap, respectively. The absorbance coefficient $\alpha(h\nu)$ and band gap $E_g$ was calculated by formula (2) and (3) as following.

$$\alpha(h\nu) = \frac{2.303A}{l} \quad (2)$$

$$E_g = \frac{hc}{\lambda} \quad (3)$$

where $l$ was the thickness of the sample. A was the absorbance. $h$ was Planck's constant. $c$ was the speed of light. $\lambda$ was the related wavelength. Plots of $[\alpha(h\nu)]^{1/2}$ and $\alpha(h\nu)^2$ versus $E_g$ were drawn according to the equations. The optical band gap was evaluated by extrapolation of the linear portion near the onset of absorption edge of the energy axis, as seen in Fig.3. The structural modification such as the formation of defects and amorphous was an important factor which can affect their physical properties. Urbach energy analysis was carried out in present work to compare the degree of defects and amorphonization in pristine and irradiated mica. It can be calculated according to the Urbach rule [22]:

$$\alpha(h\nu, T) = \alpha_0 \exp\left(\frac{\sigma h\nu - E_0}{kT}\right) \quad (4)$$

where $\alpha_0$, σ and $E_0$ were the parameters related to material properties. The logarithm of equation (4) was given below:

$$\ln\alpha = \frac{h\nu}{E_{Ur}} + C \quad (5)$$

here, $E_{Ur}$ was the Urbach energy. The dependence of logarithm of α on photon energy of pristine and irradiated mica was depicted in Fig.3(c) and (f). The Urbach energy $E_{Ur}$ can be calculated by taking the reciprocal of the slopes of the linear portion of the plots of ln (α) vs. $hv$. The values of the band gap and Urbach energy of pristine and irradiated mica were shown in Table 2.

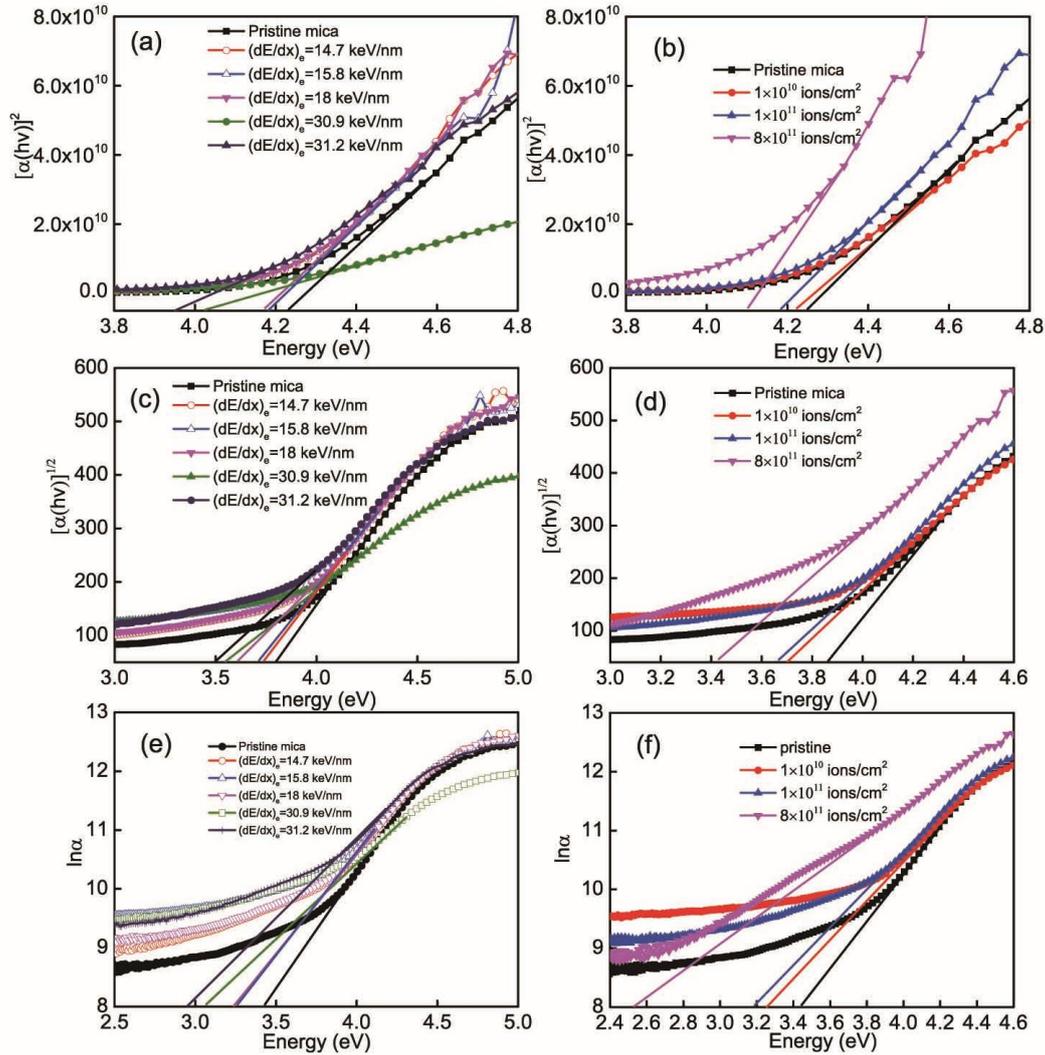

Fig.3 Varitions of band gap and Urbach energy of mica irradiated under different parameters. (a), (c), (e) were the direct, indirect band gap and Urbach energy of mica irradiated under different $(dE/dx)_e$; (b), (d), (f) were the direct, indirect band gap and Urbach energy of mica irradiated under different fluences.

Table 2 Variations of band gap and Urbach energy of mica irradiated under different parameters.

| Fluence/(ions/cm$^2$) | $(dE/dx)_e$/(keV/nm) | Direct band gap /(eV) | Indirect band gap /(eV) | Urbach energy/(eV) |
|---|---|---|---|---|
| Pristine mica | 0 | 4.19 | 3.79 | 0.29 |
| 1×10$^{11}$ | 14.7 | 4.14 | 3.73 | 0.31 |
| 1×10$^{11}$ | 15.8 | 4.14 | 3.72 | 0.31 |
| 1×10$^{11}$ | 18 | 4.13 | 3.56 | 0.32 |
| 1×10$^{11}$ | 30.9 | 3.90 | 3.55 | 0.33 |
| 1×10$^{11}$ | 31.2 | 3.85 | 3.51 | 0.34 |
| 1×10$^{10}$ | 18 | 4.17 | 3.61 | 0.31 |
| 8×10$^{11}$ | 18 | 4.07 | 3.33 | 0.40 |

Optical band gap was closely related to the material internal defects, impurities and structure. The Urbach energy parameter $E_{Ur}$ was an indicative of disorder of the material. SHIs irradiation can lead to the formation of defects, structural damage, or even amorphous transformations. Since the energy of SHIs was deposited along the path, the atomic displacement that resulted in the structural modifications of studied material would be emerged. Concluded from Fig.3 and Table 2, the direct and indirect band gap of irradiated mica were gradually narrowed down with the increasing (dE/dx)$_e$ and fluences, while the values of Urbach energy were enlarged in reverse. When (dE/dx)$_e$ was 15.8 keV/nm, the direct band gap and indirect band gap decreased from 4.19 eV and 3.79 eV to 4.14 eV and 3.73 eV, respectively. While as the electronic stopping loss (dE/dx)$_e$ increased to 31.2 keV/nm, the values of direct and indirect band gap were reduced to 3.85 eV and 3.51 eV. Meanwhile, values of Urbach energy were increased from 0.29 eV to 0.34 eV with the (dE/dx)$_e$ increased to 31.2 keV/nm. This phenomenon indicated that the formation of defects and amorphous crystal structure became much more obvious with the increasing (dE/dx)$_e$, which had been confirmed by many other investigate methods [8, 23]. The increase in the amorphous fraction resulted in the increase of the Urbach energy.

V.Chailley et al [24] found that the proportion of amorphous structure was increased with the increasing fluences. As described in Fig. 3 and Table 2, the direct and indirect band gap decreased to 4.07 eV and 3.33 eV, respectively, as ions fluences increased from 0 to $8\times10^{11}$ ions/cm$^2$ when (dE/dx)$_e$ was 18 keV/nm, the Urbach energy was increased to 0.40 eV. This increased value of Urbach energy of irradiation mica may be attributed to the increase of the amorphous crystal structure with increasing fluences. It was consistent with the previous investigation.

The values of band gap and Urbach energy of mica irradiated under different parameters were described in Fig.4. Zhu et al also found the reduction of band gaps in nc-Si:H films irradiated by 65 MeV Kr-ions under different ion fluences[18]. The values of band gap and Urbach energy were sensitive to the fluences. During the slowing down process of swift heavy ions, the energy was transformed into ejected target electrons. Most of the primary ionizations and excitations occurred close to the ion trajectory in a cylindrical region with a radius of a few nanometers called track core. Amorphous was the most well-known phase transformation in the core [25]. It was the track-halo around the core. The ionizations in the track halo were mainly caused by secondary collision. The structure in the halo was considered to be dilated crystalline. Mean diameters of the latent tracks irradiated under different (d$E$/d$x$)$_e$ were increased in a form of a function with (d$E$/d$x$)$_e$ in previous work [24]. Hence, the proportion of amorphous structure was increased at the same time. Nevertheless, the quantity of latent tracks was increasing with the increasing fluences. Damaged regions would be overlapped as the fluences reached to a particular point. The overlapping of latent tracks would give a rise to the concentration of the amorphous structure. It appeared as a collective effect for band gap and Urbach energy. Consequently, the band gap narrowed down and the Urbach energy increased with the increasing fluences.

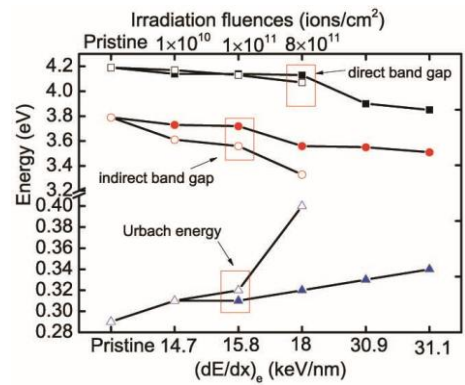

Fig. 4 Variations of optical band gap energy and Urbach energy in mica irradiated under different fluences and (dE/dx)$_e$. (■● ▲ the direct, indirect band gap and Urbach energy under different (dE/dx)$_e$, respectively; □○△ the direct, indirect band gap and Urbach energy under different fluences, respectively.)

4. Conclusion

In the present experiment, UV-VIS spectroscopy was applied to investigate the modification of mica optical properties as the samples were irradiated by SHIs with different parameters. The band gaps and

Urbach energy of pristine and irradiated mica were calculated according to the UV-VIS spectra. With the increasing of the values of $(dE/dx)_e$, the regular sine wave in the long wavelength was disturbed. It was indicated that there was cleavage appeared on the irradiated samples surface. A red shift was found in the UV-VIS spectroscopy of mica. The band gaps were narrowed down with the increasing $(dE/dx)_e$ values and ions fluences. The values of Urbach energy were increased as the $(dE/dx)_e$ values and fluences gradually increased. It was suggested that the chemical bonds were broken and the perfect structure of mica was modified due to the SHIs irradiation. Consequently, the amount of defects and the proportion of amorphous structure were increased with the increase of the $(dE/dx)_e$ values and ions fluences. The ions fluences had a great effect on the amount of defects and the proportion of amorphous.